\documentstyle[twocolumn,aps,epsf,amsbsy]{revtex}

\newcommand{\Past}		{ \stackrel{\leftarrow} {S} }
\newcommand{\past}		{ {\stackrel{\leftarrow} {s}} }
\newcommand{\Future}		{ \stackrel{\rightarrow}{S} }

\newcommand{\FutureL}		{ {\stackrel{\rightarrow}{S}}^L }

\newcommand{\CausalState}               { {\cal S} }

\newcommand{\CausalStateSet}            { \boldsymbol{\CausalState} }

\newcommand{\Prob}                              { {\rm Pr} }

\begin{document}
\bibliographystyle{unsrt}

\title{Information Bottlenecks, Causal States, and Statistical Relevance
Bases:\\
How to Represent Relevant Information in Memoryless Transduction}
\author{Cosma Rohilla Shalizi\thanks{Permanent address: Physics
Department, University of Wisconsin, Madison, WI 53706.}
and James P. Crutchfield}
\address{Santa Fe Institute, 1399 Hyde Park Road, Santa Fe, NM 87501\\
Electronic address: \{shalizi,chaos\}@santafe.edu}

\date{15 June 2000}
\maketitle
\begin{abstract}
Discovering relevant, but possibly hidden, variables is a key step in
constructing useful and predictive theories about the natural world.
This brief note explains the connections between three approaches to
this problem: the recently introduced information-bottleneck method,
the computational mechanics approach to inferring optimal models,
and Salmon's statistical relevance basis.

\end{abstract}


\section{Introduction}

Recently, Tishby, Pereira, and Bialek proposed a new method for finding
concise representations of the information one set of variables contains
about another \cite{Tishby-Pereira-Bialek-bottleneck}. This brief note
explains the connections between this ``information-bottleneck'' method
and existing mathematical frameworks and techniques. This comparison should
enhance the value of research in this promising direction and clarify the
relative uses of these techniques in applications.

In the interest of space, we assume readers are familiar with the notation
of both \cite{Tishby-Pereira-Bialek-bottleneck} and \cite{TDCS}.

\section{The Information-Bottleneck Method}

In \cite{Tishby-Pereira-Bialek-bottleneck} the authors pose the following
problem.  Given a joint distribution over two random variables---the
``input'' $X$ and the ``output'' $Y$, find an intermediate or ``bottleneck''
variable $\tilde{X}$ which is a (possibly stochastic) function of $X$ such
that $\tilde{X}$ is more compressed than $X$, but retains predictive
information about $Y$.  More exactly, they ask for a conditional distribution
$\Prob (\tilde{x}|x)$ that minimizes the functional
\begin{equation}
{\cal F} = I[\tilde{X};X] - \beta I[\tilde{X};Y] ~,
\end{equation}
where $I[W,Z]$ is the mutual information between random variables $W$ and $Z$
\cite{Cover-and-Thomas} and $\beta$ is a positive real number. Minimizing the
first term
represents the desire to find a compression of the original input data $X$;
maximizing the second term represents the desire to retain the ability to
predict $Y$.\footnote{Since $\tilde{X} = g(X, \Omega)$ for some auxiliary
random variable $\Omega$, a theorem of Shannon's assures us that
$I[\tilde{X}; Y] \leq I[X; Y]$ and the transformation from $X$ to $\tilde{X}$
cannot {\it increase} our ability to predict $Y$ \cite[App. 7]{Shannon-1948}.}
The coefficient $\beta$ governs the trade-off between these two goals: as
$\beta \rightarrow 0$, we lose interest in prediction in favor of compression;
whereas as $\beta \rightarrow \infty$, predictive ability becomes paramount.

Extending classical rate-distortion theory, the authors are not only able to
state self-consistent equations that determine which distributions satisfy this
variational problem, but give a convergent iterative procedure that finds one
of these distributions. They do not address the rate of convergence.

\section{Causal States for Transducer-Functionals}

In \cite{TDCS} and earlier publications, we defined causal states
for stationary stochastic processes as follows. Two histories $\past$ and
${\past}^{\prime}$ belong to the same causal state $\CausalState$ if and
only if
\begin{equation}
\Prob(\FutureL = s^L|\Past = \past)
  = \Prob(\FutureL = s^L|\Past = {\past}^{\prime}) ~,
\label{CausalStateEqReln}
\end{equation}
for all $s^L$ and for all $L$. That is, two histories belong to the same
causal state if and only if they give the same conditional distribution for
futures. In \cite{TDCS} we showed that the causal states defined by Eq.
(\ref{CausalStateEqReln}) possess two kinds of optimality. First, their
ability to predict the future $\Future$ is maximal. Second, they are the
simplest such set of states.

Thus, we showed that the set $\CausalStateSet$ of
causal states is the solution to the following optimization problem. Given
the joint distribution $\Prob(\Past, \Future)$ over the past $\Past$ and
future $\Future$, find the function (equivalently, partition) $\epsilon$
of $\Past$ such that (i) the conditional entropy $H[\FutureL|\epsilon(\Past)]$
is minimized for all $L$, and (ii) the entropy $H[\epsilon(\Past)]$ is
minimized among all functions $\hat{\eta}$ that satisfy condition
(i).\footnote{The states induced by the partition $\hat{\eta}$ are called the
prescient rivals of the causal states induced by $\epsilon$. The entropy
$H[\epsilon(\Past)]$ is called the statistical complexity and measures the
``size'' of, or amount of information stored in, the causal states.} The
equivalence classes induced by $\epsilon$ are the causal states, and they
are the unique\footnote{More precisely, any other function satisfying conditions
(i) and (ii) may differ from $\epsilon$ on at most a set of histories of
measure zero.} solution to the optimization problem. A moment's reflection
shows that this optimization is equivalent to first maximizing the mutual
information between the effective states and the futures and then minimizing
the mutual information between the effective states and the histories. The
first step maximizes the predictive ability of the effective states and the
second selects the most concise set of states. Note that the opposite sequence
of optimizations---first minimizing complexity and then maximizing
predictability---is trivial, since it produces a single-state model that
describes an IID sequence of random variables; e.g., a biased coin or die.

\section{Memoryless Transducers}

As has been remarked earlier---e.g., \cite{Inferring-stat-compl} and
\cite{Calculi-of-emergence}---the
causal-state construction is not intrinsically limited to time series. Of
particular interest here is the case of transducer-functionals: when one
sequence of variables is a functional of another sequence, possibly a
stochastic functional. In this case, one can construct causal states that
(i) retain all predictive information about the output series, (ii) are
deterministic functions of the prior causal state and the most recent value
of the input series, and (iii) minimize the statistical complexity of
(information stored in) the causal states. We present the theory of causal
states for general transducers elsewhere.

In the case where the output depends on the {\it current} input alone---the
case of memoryless transduction---the causal states assume a particularly
simple form: two inputs $x$ and ${x}^{\prime}$ belong to the same causal
state if and only if
\begin{equation}
\Prob(Y = y| X = x) = \Prob(Y = y| X = {x}^{\prime}) ~,
\end{equation}
for all $y$.\footnote{A somewhat more complicated construction is necessary
when the transducer exhibits memory.} In this case, it can be shown that
\begin{equation}
H[Y|\epsilon(X)] \leq H[Y|\eta(X)] ~,
\end{equation}
for any other partition of
the inputs $\eta$ and that, among all the rival partitions $\hat{\eta}$
minimizing the conditional entropy of the outputs (the prescient rivals of
$\epsilon$),
\begin{equation}
H[\epsilon(X)] \leq H[\hat{\eta}(X)] ~.
\end{equation}
This is to say, the causal states are the most compressed hidden variables.
In the sense of \cite{Tishby-Pereira-Bialek-bottleneck}, they are optimal
bottleneck variables.

One concludes that these are precisely what should be delivered by the
information-bottleneck method in the limit where $\beta \rightarrow \infty$.
It is not immediately obvious that the iterative procedure of
\cite{Tishby-Pereira-Bialek-bottleneck} is still valid in this limit.
Nonetheless, that $\epsilon$ is the partition satisfying their original
constraints is evident.

We note in passing that, as shown in \cite{TDCS}, prescient rivals---those
sets of states that retain all predictive information from the original inputs
while compressing them---are sufficient statistics. Conversely,
\cite{Tishby-Pereira-Bialek-bottleneck} states that, when sufficient
statistics exist, then compression-with-prediction is possible.

\section{The Statistical Relevance Basis}

Before closing, we point out another solution to the problem of discovering
concise and predictive hidden variables. In his books \cite{Salmon-1971} and
\cite{Salmon-1984}, Salmon put forward a construction, under the name of the
``statistical relevance basis'', that is identical in its essentials with that
of causal states for memoryless transducers.\footnote{Salmon's work only came
to our attention in mid-1998, and thus it is not cited in our publications
including and prior to \cite{TDCS}.}  Owing to the rather different aims for
which Salmon's construction was intended---explicating the notion of
``causation'' in the philosophy of science, no one seems to have proved its
information-theoretic optimality properties nor even to have noted its
connection to sufficient statistics. (Briefly: if a nontrivial sufficient
partition of the input variables exists, then the relevance basis is the
minimal sufficient partition. These proofs will appear elsewhere.)

\section{Comparison and Conclusion}

Recapitulating, the causal states for memoryless transduction coincide with the
cells of the ``bottleneck'' partition in the limit $\beta \rightarrow \infty$.
Moreover, both are identical with the statistical relevance basis of Salmon.

The construction of the causal states does not allow us to discard {\it any}
predictive information about the output $Y$, even if this might allow for a
substantial reduction in the statistical complexity.  The bottleneck method, by
contrast, generally throws away {\it some} predictive information. It trades
one bit less statistical complexity for $1 / \beta$ bits less predictive
information. Of course, the linear trade-off and the particular value of the
coefficient $\beta$ that controls it are {\it ad hoc} choices. Whether this is
acceptable in applications would seem to depend on the goal. For example, if
the goal is a practical ``lossy'' data-compression scheme, the bottleneck
method recommends itself. However, if the goal is representing the intrinsic
computation or causal structure of some natural process, causal states are
better suited to the task.

\section*{Acknowledgments}

This work was partially supported by the SFI Computation, Dynamics, and
Learning Program, by AFOSR via NSF grant PHY-9970158, and by DARPA under
contract F30602-00-2-0583.

\bibliography{locusts}

\begin{thebibliography}{1}

\bibitem{Tishby-Pereira-Bialek-bottleneck}
Naftali Tishby, Fernando~C. Pereira, and William Bialek.
\newblock The information bottleneck method.
\newblock Electronic pre-print, LANL archive, physics/0004057, 2000.

\bibitem{TDCS}
James~P. Crutchfield and Cosma~Rohilla Shalizi.
\newblock Thermodynamic depth of causal states: {O}bjective complexity via
  minimal representations.
\newblock {\em Physical Review E}, 59:275--283, 1999.

\bibitem{Cover-and-Thomas}
T.~M. Cover and J.~A. Thomas.
\newblock {\em Elements of Information Theory}.
\newblock Wiley, New York, 1991.

\bibitem{Shannon-1948}
Claude~E. Shannon.
\newblock A mathematical theory of communication.
\newblock {\em Bell System Technical Journal}, 27:379--423, 1948.

\bibitem{Inferring-stat-compl}
James~P. Crutchfield and Karl Young.
\newblock Inferring statistical complexity.
\newblock {\em Physical Review Letters}, 63:105--108, 1989.

\bibitem{Calculi-of-emergence}
James~P. Crutchfield.
\newblock The calculi of emergence: {C}omputation, dynamics, and induction.
\newblock {\em Physica D}, 75:11--54, 1994.
\newblock http://www.santafe.edu/projects/CompMech.

\bibitem{Salmon-1971}
Wesley~C. Salmon.
\newblock {\em Statistical Explanation and Statistical Relevance}.
\newblock University of Pittsburgh Press, Pittsburgh, 1971.
\newblock With contributions by Richard C. Jeffrey and James G. Greeno.

\bibitem{Salmon-1984}
Wesley~C. Salmon.
\newblock {\em Scientific Explanation and the Causal Structure of the World}.
\newblock Princeton University Press, Princeton, 1984.

\end{thebibliography}

\end{document}